English language: *no proof read*.

**Title**:

Suppression and facilitation of motion perception in humans: a reply to Schallmo & Murray (2018)

**Authors**: Tzvetomir Tzvetanov

**Affiliation**: Anhui Province Key Laboratory of Affective Computing and Advanced Intelligent Machine, School of Computer Science and Information Engineering, Hefei University of Technology, Hefei, P.R. China

**Contact**: tzvetan@hfut.edu.cn

# Abstract

In a recent publication (Tzvetanov (2018), bioRxiv 465807), I made an extensive analysis with computational modelling and psychophysics of the simple experimental design of Dr. D.Tadin (Tadin, Lappin, Gilroy and Blake (2003), Nature, 424:312-315) about motion perception changes in humans due to size and contrast of the stimulus. This publication sparked from strong claims made in Schallmo et al. (2018) (eLife, 7:e30334) about two important points: (1) "divisive normalization", not inhibitory and excitatory mechanisms, creates the observed psychophysical results and (2) drug-enhanced inhibition showed perceptual outcomes that hint to "weaker suppression" (i.e. inhibition) not stronger "suppression". Schallmo & Murray (2018, bioRxiv, 495291) presented concerns about my extensive publication, specifically about the parts where I directly analysed some of their methods, results and claims. Here, I show that their concerns do not provide clear answers to my specific points and further do not mention other major critiques of data interpretation and modelling of this experimental design. Therefore, I maintain all my claims that were elaborated in details in my first publication (Tzvetanov, 2018, bioRxiv 465807): the specific ones that analyse the results of their and other studies, but also the more broad modelling that is applicable to any study using the simple experimental design of Dr. D.Tadin.

## Brief introduction

Tadin et al. (2003) demonstrated an interesting and counter-intuitive outcome on motion perception of a very simple moving stimulus. In order to correctly report the direction of motion of a unidirectional moving pattern (e.g. Gabor patch, random dot kinematogram; motions only in two possible opposite directions), when increasing the size of the stimulus it would be natural that the capacity of seeing the motion direction is improved. They showed that this happens when the stimulus has low luminance contrast, by demonstrating that observers needed much less stimulus presentation time for larger sizes in order to correctly discriminate the direction of motion. Interestingly, the opposite was present when stimulus contrast was very strong. Thus, it was proposed that perception of such a simple stimulus allows to extract knowledge about excitatory-inhibitory spatial interactions due to contrast and size that are typical of center-surround motion tuned neurons in area MT/V5.

Later studies (Tadin & Lappin, 2005; Betts, Sekuler & Bennett, 2012; Schallmo et al., 2018)  proposed and used a computational model in order to explain in details the perceptual results. The latest report of Schallmo et al. (2018) made some strong claims from the data and modelling.

In my report Tzvetanov (2018), I analysed the global experimental design of Dr. D.Tadin from psychophysics and modelling point of view. Multiple points were presented, some of which directly related to the psychophysics methods, computational modelling, and data inferences presented in Schallmo et al. (2018). The conclusions I drew from modelling and some psychophysics data reported by Schallmo et al. (2018) were different from the claims in their paper.

Schallmo and Murray (2018) took my extensive analyses as a simple commentary of Schallmo et al. (2018). They summarized it in two points (from their abstract): (1) "response bias" and (2) I presented an "other computational models".

In this reply, I provide answers to the points put forward by Schallmo et al. (2018).

## General comment

Concerning the psychophysics part. The reason I presented (Tzvetanov, 2018) in such details the definition of the psychometric function in Dr. D.Tadin's design was that it is important to properly define the psychometric function in order to link it to the computational model that is used to predict the data, and finally to propose how perception is shaped by neurophysiological knowledge. The use in this design by Schallmo et al. (2018) of the Weibull ad-hoc model of psychometric function together with pooled left-/rightward motion directions in the data analysis are creating unnecessary complications in threshold estimates (biases) and interpretation with respect to their proposed model. The reply by Drs Schallmo and Murray (2018) does not acknowledge this important point. Instead, it discusses whether psychometric function shift can be estimated from their data and how it could be implemented.

Additionally, they do not reply to a major critique of the use of "Size Index" as labelling "suppression" and "facilitation" effects, which I clearly argued can not be used in their study.

Concerning the Modelling part. In the Modelling section of Tzvetanov (2018), I presented seven subsections that analysed different parts of the modelling approach in this experimental design, I derived the correct model, I extended the correct model to quantitative fitting to the data (methods I made openly available), and I analysed how the combination of psychophysics results and model fits predict about the underlying neurophysiological changes. Schallmo and Murray (2018) sum-marized these 9 pages of model analyses and data fits as if it was "*a different computational framework from the normalization model of Reynolds and Heeger (2009) provides a better description of the direction discrimination psychophysical data*" (their first sentence of Introduction to issue #2).

Here I repeat the titles of the seven subsections in the Modelling of Tzvetanov (2018):

- "A note on normalization of neuronal activity"
- "Concerns in Schallmo et al. (2018)'s model of motion perception and its application"
- "General background for the modelling"
- "The correct model for D. Tadin's design"
- "Comparing predictions of the two low-level models"
- "Application of the model"
- "Conclusion from modelling"

Thus, my presentation is not a different computational model. I described what I considered to be the correct model for Dr. D.Tadin's experimental design and I claimed that the model used in the study of Schallmo et al. (2018), despite that its final mathematical equation provides description of the threshold data, is not based on sound computational and psychophysical modelling principles (specifically the problem of decision stage and "Criterion" in their model).

Below are detailed commentaries to their reply.

## Psychophysics

In the Results section of Tzvetanov (2018) about Psychophysics I presented two subsections discussing the data analyses and presentation used by Schallmo et al. (2018) : (1) the biased measures of thresholds that are present in the analysis of the design as it was used and (2) the variable "Size Index" they reported cannot be used for labelling "suppressive" and "facilitative" effects in this paradigm of motion perception given the data results they reported. The reply by Schallmo and Murray (2018) only targets the first point – they discussed whether it is possible to estimate "response bias" instead of directly reporting the unbiased threshold/slope of the psychometric function. It does no mention the issue about "Size Index" and the labels "suppression" and "facilitation". Both points are re-commented and replied below.

### On the issue of "Size Index" (SI) and definition of "suppression" and "facilitation" of motion perception in humans

I note that Schallmo and Murray (2018) did not reply to my critique of the use of "Size Index" (SI) as a variable to define what is "suppression" and "facilitation" of motion perception. In their reply they write about "suppression" and "facilitation" without clearly defining their meaning in the reply, thus assuming the same meaning as in Schallmo et al. (2018). Thus, they report "Size Index" results as "suppression" (SI<0) and "facilitation" (SI>0) despite of clear evidence in their own data set of the difficulty to use it as such. One important point is that "suppression" and "facilitation" are not clear psychophysical concepts in the article of Schallmo et al. (2018) and cannot be used to label "inhibition" and "excitation" as it was done (Schallmo et al. (2018) , page 6: "*We found that increasing inhibition via lorazepam did not lead to stronger suppression – in fact, we observed weaker spatial suppression (SIs were less negative overall)*"; SI stands for "Size Index").

**Bias on duration threshold estimates due to psychometric function midpoint shifts because of "response biases"**

From the arguments put forward by Schallmo and Murray (2018), it seems there is a misunderstanding about the importance of biased threshold (proportional to inverse of slope) estimates due to psychometric function shifts from "zero" ("response biases"), what to expect on bias variability when fitting the psychometric function to the data, together with how to extract the subjects' response bias from their data.

Schallmo and Murray (2018) claim that I have shown the "response bias" to be a small shift on the x-axis (page 1, paragraph 2: "*As shown by Tzvetanov (2018) (Figure 2A), this is equivalent to a small shift of the psychometric function along the x-axis.*"). Figure 2A in Tzvetanov (2018) shows a hypothetical midpoint shift of one threshold value given that the theoretical threshold has a value of one. The point of this example was not to demonstrate a "strong" or "small" shift but to show how the method of analysis applied by Schallmo et al. (2018) leads to overestimates of the theoretical threshold (slope). With their analysis method the estimated thresholds depend simultaneously on both midpoint (shift from zero) and real threshold (slope) of the function. For the particular example that was given, psychometric function shift of midpoint of one threshold value leads to threshold overestimation of ~52% when applying the analysis procedure of Schallmo et al. (2018) (page 4, paragraph 2 in Tzvetanov (2018)).

Whether the final experimental estimates of thresholds have large or small biases is hard to know in advance unless one analyses all the data to check for it. I also note that the example fit provided by Schallmo and Murray (2018) in their Figure 1 does not compare correctly when the data are processed with their method (pooling left and rightward motions) and when considering the full range of left- and rightward motions, because both data sets are not fit with the same theoretical psychometric function. Instead they fit the different mathematical models to the different data processing methods, which does not give a correct comparison. It would have been also more appropriate, instead of presenting a data set with mostly extremely low threshold estimates (~20ms), to have shown a data set result within the common thresholds they reported across their experiments (range of mean values across Figures 2C, 4A-B, and 5A-B in Schallmo et al. (2018) : ~30ms to ~150ms, with majority within 50-100 ms range).

Schallmo and Murray (2018) claim that their data does not allow to extract reliable fits with midpoint ("bias") values within ±5ms (page 1, 4[th] paragraph: "*A low number of trials (< 5) at very short stimulus durations, and unequal sampling of left and right trials, often made it unfeasible to obtain reliable fits to a Logistic model function with threshold values (reflecting the absolute response bias) in the expected range of ±5 ms.*"). With a low number of 30 trials to measure a single condition one cannot expect that all measured midpoint values are within ±5 ms, even in the perfect case of theoretical no shift. It depends on the number of trials but also on the theoretical psychometric function slope (shallower functions will give larger variations of the midpoint across measures) and thus midpoint of the psychometric function should be expected to be more variable than the ±5 ms limit they constrained themselves, depending on the slope of the function. Furthermore, they put forward the fact that their design did not measure too many responses to small drifting times (0 or 1/85 or 1/120 s). This is not astonishing, neither so important for the fitting procedure.

The question is how to extract the bias in the design they used, such that one can obtain the best unbiased threshold estimates. Instead of adding supplementary trials at small drifting times, as they argue, and indeed one can spend experimental time to do it, the data they have collected for each subject can be readily used to extract each individual observer midpoint within the block of measure. And this is not a "*post-hoc*" analysis, as they suppose it, but an analysis coming from the constrains of their experimental design.

Two simple methods can be used. The first one is to check whether the 6 estimated midpoints of a given observer have some systematic deviation. While simple, this method has the inconvenience that when using low number of trials for psychometric function estimate the two psychometric function parameters are not very well constrained. Therefore, a better approach is the second method as follows. One must not fit each condition separately with 2 parameters for each psychometric function, {*midpoint*, *threshold/slope*}, which makes the maximum-likelihood fits become frequently unreliable with small number of trials. Because of the design, where all conditions were measured within a single block, one could assume that bias is equally affecting all conditions and thus instead of their 6*2=12 free parameters one has 6+1 free parameters, with six different thresholds left free to vary and one common bias value across all conditions. This assumes that there are no particular conditions that induce different perceptual/response biases of the observers in comparison to the other ones. (remarque: one can further assume the lapse rate is the same across conditions, thus allowing for different lapse rates across observers, instead of fixing it at the same value for all observers). Last, the authors used the maximum-likelihood method with fixed lapse rate for all subjects and conditions, and thus no constrains on the thresholds/slopes. With low-number of trials used for estimating psychometric function, it is advisable to use Bayesian fitting in order to constrain the slope in some a priori known range. This constrain helps avoiding some unrealistic fit outcomes, as they have found and properly eliminated in their study when using the Weibull ad-hoc model. It can be combined with lapse rate Bayesian constrain, thus allowing a different data handling that should bring a better data analysis.

The above fitting method is not necessary to be implemented in an adaptive procedure, as proposed by Schallmo and Murray (2018) (page 2, 2nd paragraph), although it should not give special technical difficulties to do it.

Last, Schallmo and Murray (2018) seemed to keep considering the Weibull model defined between 50%-100% for the psychometric function more appropriate than the model defined between 0%-100% (Figure 2A in Tzvetanov (2018) presents a cumulative Gaussian function, not a logistic function). But in this design the Weibull model has mainly disadvantages. The main reasons are that it uses two parameters for estimating the threshold of the observers, the stimulus value giving a fixed amount of percent "correct", and especially how to relate this threshold to the model prediction. Using two parameters to estimate a single one is one too many, especially when the design and method of analysis allows to avoid it. Contrary to the Weibull model, the cumulative Gaussian model or the logistic model are theoretically justified since they are directly related to the theoretical prediction of the computational model, respectively SDT or DDM (Equation 10 in Tzvetanov (2018)).

Despite of biased threshold estimates that come from inadvertently not taking into account psychometric function shifts, I acknowledged already in Tzvetanov (2018) (page 12, lines 452-454) that, because of the overall design of the study, the biases of thresholds can be considered as a constant factor across conditions of a single observer. The overall thresholds can be considered as being a scaled version of the original theoretical value (when plotting thresholds versus stimulus size, a shift of the data along the ordinate). The important consequence is that the computational model can still be used on the extracted thresholds because the bias is readily taken care of by a single parameter in the model, and thus does not affect the estimates of the other computational model parameters that are the most interesting ones.

## Modelling

As it was stated in the "General comment" at the beginning of this manuscript, Schallmo and Murray (2018) regard my model presentation as a different model.

I clearly presented in the modelling section of Tzvetanov (2018) that when modelling psychophysics results with the general framework of "pattern analyzers", each computational model is split into two stages: a low-level response of neuronal population and a high-level stage using the low-level responses to predict the psychometric function (or some derived quantity as its threshold/slope). My model presentation is first a correction of their decision stage of the model (and of other studies using this decision stage). The decision stage described in Schallmo et al. (2018) is contradictory to common psychophysics and computational modelling knowledge. I acknowledge that the low-level model in Tzvetanov (2018) is different from the one in Schallmo et al. (2018). This difference comes also from strong concerns about their low-level model, concerns not explicitly stated in Tzvetanov (2018), though they can be clearly inferred from subsections "General background for the modelling" and "The correct model for D. Tadin's design". In their model, there are difficulties to associate all independent variables used of contrast, size, and duration of stimulus to different low-level components of the model that code these variables. Specifically, (1) contrast seemed to have been not modelled, but instead put at the "decision level" stage in the variable "Criterion" (in Schallmo et al. (2018), page 23, Appendix 1-Table 1, column "Basic model"), and (2) the variable "stimulus duration" was also not introduced in the low-level stage of the model but instead they argued to come from the "decision stage" level, somehow related to the "drift-diffusion model" of decision making. In Tzvetanov (2018) I wrote "*their model relating perceptual Threshold and low-level model neuronal responses (Equation 6) is a new construct whose origin and derivation is not clearly given*". Schallmo and Murray (2018) did not comment on this. I also acknowledged that the decision stage model was borrowed from previous publications that also did not clearly present the origin and derivation of the final equation (in Tzvetanov (2018), page 7, 1[st] paragraph, "*The exact model description of how this equation is obtained is not fully described in Schallmo et al. (2018), neither in the two other studies using it (see Appendix A2)*"). Thus, my publication (Tzvetanov, 2018) presents a large extension and correction of the modelling and its application.

All the remaining text of Schallmo and Murray (2018) about the modelling presents claims, which after inspection some of them turn out to be very weak.

For instance, in their "Introduction to issue #2 – computational modelling" :

- they summarize the modelling presented in Tzvetanov (2018) by claiming that a different model provides a better description of the data than the "normalization model" they used (1[st] sentence: "[...]*a different computational framework from the normalization model of Reynolds and Heeger (2009) provides a better description of the direction discrimination psychophysical data*"). On the contrary, I demonstrated that two models of low-level neuronal activity, the "normalization model" and the "subtractive model", provide exactly the same mathematical outcome and they cannot be dissociated in the experimental design of D.Tadin (subsection "Comparing predictions of the two low-level models"), while the decision stage model in their publication is not substantiated.

- they claim that "*We adopted the Reynolds and Heeger (2009) normalization model in an effort to: [...] 2) determine whether normalization could account for spatial summation (in addition to suppression),*"; "divisive normalization" was used in the report of Betts et al. (2012) (see their Figure 1, 3[rd] row of panels and equations, where additionally contrast coding is clearly modelled), and thus already demonstrated to "*account for spatial summation (in addition to suppession)*", though with a wrong "decision stage" model.

- they claim that "*We adopted the Reynolds and Heeger (2009) normalization model in an effort to [...] 3) establish a model framework within which we could more easily interpret the* *results from our experiments using lorazepam and MR spectroscopy*" (1ˢᵗ sentence, page 3 in Schallmo and Murray (2018)); I demonstrated that "*their model and its application contains conflicting knowledge with established computational neuroscience and behavioural modelling*" (page 2, lines 77-78 in Tzvetanov (2018)), thus it cannot be used to interpret the data. Once the model is correctly established, together with the threshold data correctly interpreted (not the "Size Index"), it turns out that one important interpretation that was made in Schallmo et al. (2018) about the effect of Lorazepam is opposite to what the combined data and correct model interpretation gives.

In their "Discussion – computational modeling":

- In the first paragraph, they write that I misunderstood their claims. The quote from my publication they showed is extracted from this full sentence:

    ○ "*Thus, they proposed that the overall effects and observations for the single grating design of D. Tadin and collaborators (and other researchers who use it) are <u>not necessarily due to excitatory and inhibitory interactions between motion sensitive neurons but instead arises from the divisive normalization of neuronal activity</u>, that emerges from classic neuronal network computations, and which they, and others, have used as a "computational principle"/"computational framework".*" (Tzvetanov (2018), page 2, lines 66-71); they quoted the underlined part.

    In the Abstract of Schallmo et al. (2018) it is written:

    ○ "*We show that: (1) both suppression and facilitation can emerge from a single, computational principle – divisive normalization; there is no need to invoke separate neural mechanisms*"

    This is a clear statement in their article, unless they state explicitly that "*separate neural mechanisms*" do not relate to the separate mechanisms of inhibition and excitation.

- In the 2ⁿᵈ paragraph, they discuss about "*the relationship between the behavioral phenomenon of spatial suppression and the underlying neural processes*"; "spatial suppression" is not clearly present in their data and thus not clearly defined in the psychophysics results they presented. See above, section "**On the issue of "Size Index" (SI) and definition of suppression and facilitation of motion perception in humans**". I claimed that "[...]*contrary to the reported composite "Size Index", the threshold data should be used directly to infer something about changes of inhibition and excitation in the visual motion system of the observers. This last point will appear very clearly once the model is established and used to predict how inhibition and excitation, together, shape the thresholds*." (in Tzvetanov (2018), page 5, lines 176-179).

- In the 3ʳᵈ paragraph, they state : "*After presenting an alternative computational model to the divisive normalization framework, Tzvetanov (2018) goes on to critique our interpretation of our lorazepam data in light of the newly proposed model results.*"

    I demonstrated that their "decision stage" model is not based on sound computational and psychophysical modelling principles. I demonstrated that "divisive normalization" and "simple subtraction", two competing models of low-level activity, provide exactly the same mathematical prediction of thresholds and can not be dissociated in this design (in

Tzvetanov (2018), Results-Modelling, subsection "Comparing predictions of the two low-level models").

Once the model and data are correctly stated and interpreted the Lorazepam data shows psychophysics thresholds increase which are predicted by the model through inhibitory enhancement, exactly what Lorazepam is supposed to do. Schallmo and Murray (2018) continue to use log-threshold differences (Size Index) to label "suppression" and "facilitation", while I clearly have stated that "*while the exact threshold variation is dependent on the combined changes of inhibitory parameters, increase in inhibition makes overall thresholds higher*" (Tzvetanov (2018), lines 481-482).

- In the 4th and last paragraph, Schallmo and Murray (2018) write that "*The model presented by Tzvetanov (2018), which follows in the vein of Betts and colleagues (2012), is well-intended, but utilizes too many free parameters for our purposes.*". This statement shows that Schallmo and Murray (2018) did not correctly compare these published models. My model's "decision stage" is not in the vein of Betts et al. (2012). These later authors used exactly the same "decision" stage part of the model as Schallmo et al. (2018) where the threshold is obtained through a theoretically not instantiated equation, Threshold=Criterion/Response. In my publication, I have clearly stated that "*This equation is claimed to be somehow related to the DDM of decision making [...], and it seems to have been taken from the two earlier studies of Tadin and Lappin (2005); Betts, Sekuler and Bennett (2012)*" (Tzvetanov (2018), page 7, lines 262-265). Thus, I have also criticised the modelling work of Betts et al. (2012), as it can be seen in Appendix 2 of Tzvetanov (2018), but only the decision stage part. The low-level model in Betts et al. (2012) contains a "divisive normalization" model, exactly as in Schallmo et al. (2018). Thus, the their model is much in the vein of the model described in Betts et al. (2012) than the one in Tzvetanov (2018).

Schallmo and Murray (2018) write that the modelling "*utilizes too many free parameters for our purposes*". Any model of D.Tadin's design must necessarily incorporate all parameters that were clearly described in my publication. I note that the model presented in Schallmo et al. (2018) contains a total of 7 parameters (Appendix 1-Table 1, column "Basic model"), whose numerical values were found by searching for a qualitative model match to the data. Whether, for matching the model to data, parameters are free or not depends on the researchers. For quantitatively matching the correct model to their data, I demonstrated that 4 free parameters are enough, and the remaining parameters can be obtained from previously published data. It is regrettable that Schallmo and Murray (2018) seem to consider a "*good qualitative match*" (in Schallmo et al. (2018), caption of Figure 2, and Methods-Computational modeling, page 13, 2nd paragraph) of a model to data better than a quantitative match of 4 free model parameters to 6 data points. That there are only 6 data points in their data is due to the original design of the study, which it seems was not intended to provide exact quantitative model inferences. I have clearly discussed the problematic in this experimental paradigm about model parameters inference and using them to predict underlying neurophysiological parameter changes (Discussion in Tzvetanov (2018)).

## Summary

To conclude, Schallmo and Murray (2018) do not respond to major critiques of the publication of Schallmo et al. (2018). These critiques are:

- The use of "Size Index" to label "suppressive" and "facilitative" effects in the data, while instead one should directly use and interpret the raw thresholds.

- The reported perceptual thresholds contain unnecessary biases that are due to the method of analysis and that should easily be taken into account.

- Their model has a final equation relating perceptual thresholds to the computational model that is not substantiated from computational and psychophysics modelling point of view.

- The lorazepam results of higher perceptual thresholds are consistent with increased inhibition, and all the observed effects on perceptual thresholds are readily explainable through the correct model although the amount of change in each inhibitory component is hard to infer.